\documentclass[12pt]{article} 

\usepackage{natbib}
\usepackage{lineno,hyperref}

\usepackage{multirow}
\usepackage{longtable}
\usepackage{booktabs}
\usepackage{geometry}
\usepackage{graphicx}
\usepackage{setspace}      
\usepackage{amssymb}
\usepackage{amsmath,bm}
\usepackage{mathrsfs}
\usepackage{amsthm}
\usepackage{epsfig,amsmath,float}
\usepackage{graphics}
\usepackage[dvipsnames]{xcolor}
\usepackage{epsfig}

\usepackage{xr}
\usepackage{amsmath,amssymb,amsthm,bm,mathtools}
\usepackage{booktabs}
\usepackage{enumitem}
\usepackage{hyperref}
\usepackage{xcolor}
\usepackage{lmodern}
\hypersetup{colorlinks=true,linkcolor=blue!55!black,citecolor=blue!55!black,urlcolor=blue!55!black}
\numberwithin{equation}{section}
\newtheorem{theorem}{Theorem}
\newtheorem{proposition}{Proposition}

\newtheorem{remark}{Remark}
\newtheorem{assumption}{Assumption}

\DeclareMathOperator{\diag}{diag}
\DeclareMathOperator{\Diag}{Diag}
\DeclareMathOperator{\tr}{tr}

\DeclareMathOperator{\Var}{Var}

\newcommand{\Fcal}{\mathcal F_N}
\newcommand{\eps}{\varepsilon}

\let\TextOslashSmall\o
\let\TextOslashCapital\O
\def\o{\ifmmode{\scriptstyle{\mathcal{O}}}\else\TextOslashSmall\fi}
\def\O{\ifmmode\mathcal{O}\else\TextOslashCapital\fi}

\newcommand{\E}{\mathbb{E}}

\newcommand{\blind}{0}

\begin{document}

\def\spacingset#1{\renewcommand{\baselinestretch}%
{#1}\small\normalsize} \spacingset{1}

\date{}
\if0\blind
{

  \title{\bf Testing for Peer Effects in Heteroskedastic Fixed-Effects Panels}
	\author{Leheng Cai$^1$$^*$,  Weijia Li$^1$$^*$, and   Xu Guo$^2$$^\dagger$  
		\\~\\
		{\small \it $^{1}$  Department of Statistics and Data Science, Tsinghua University}
		\\{\small \it $^{2}$ School of Statistics, Beijing Normal University}
        \\{\small \it $^{*}$ The first two authors contributed equally to this article and are listed in alphabetical order.}
        \\{\small \it $^{\dagger}$ Corresponding author: xustat12@bnu.edu.cn}
	}
} 

\if1\blind
{
  \bigskip
  \bigskip
  \bigskip
  \begin{center}
    { \bf }
\end{center}
  \medskip
} \fi

\bigskip

\maketitle
	
	\textbf{Abstract:}   
 We study testing for peer effects in linear panel models with individual and time fixed effects when the interaction network is unspecified. We propose a quadratic-form test based on the Gram matrix of residualized peer covariates. A key difficulty is that fixed-effects removal and nuisance estimation destroy the centering of a conventional leave-diagonal-out quadratic form under conditional heteroskedasticity. We introduce a leverage adjustment that restores exact conditional centering and combine it with a residual-based variance estimator. The resulting studentized statistic is asymptotically standard normal as both panel dimensions diverge, without imposing any restriction on their relative growth rates.  {Under additional conditions, we establish consistency against alternatives.} The null theory is further extended to covariates that are weakly dependent over time; the same studentization remains valid without an additional long-run variance estimator. Monte Carlo experiments evaluate the finite-sample size and power of the proposed test across different settings and demonstrate its favorable performance relative to the existing benchmark test. Applications to state tax
competition and growth spillovers among OECD countries illustrate
the practical relevance in testing for peer effects without specifying the
interaction network.
 

	\maketitle
	
\newpage
\spacingset{1.5} 

\section{Introduction}\label{sec:introduction}

{
Economic outcomes often respond to the behavior and characteristics of other agents. A worker's effort may depend on the productivity of coworkers, and an economy's production may benefit from knowledge generated elsewhere. Workplace interactions and technological interdependence provide empirical settings in which such spillovers are of direct interest \citep{mas2009peers,ertur2007growth}. In these settings, a useful first question is whether outcomes can be adequately modeled without peer interactions. {Answering this question need not require recovering every link in the underlying network.} It does, however, require distinguishing the evidence for interactions from persistent individual heterogeneity and common time shocks.

Panel data provide a natural basis for this distinction, but the interaction structure is often only partially observed or entirely unavailable. For example, a panel may record outcomes and individual characteristics without recording whom each individual influences. Imposing geographic proximity, equal weights, or a common interaction coefficient then turns a test for spillovers into a joint assessment of spillovers and a particular network specification. The identification difficulties associated with simultaneous social interactions are well documented \citep{manski1993reflection,bramoulle2009identification,blume2015linear}. 
A test that does not require specifying the interaction network is therefore useful even when the magnitude and direction of individual links are not the objects of interest.

Two complications arise in this setting. First, removing individual and time fixed effects does not eliminate conditional heteroskedasticity in the idiosyncratic disturbances. Second, when the peer network is left unspecified, the number of candidate peer instruments grows on the order of \(n^2\), whereas the sample size grows only as \(nT\). Projection-based procedures that require the instrument dimension to be smaller than the sample size may therefore cease to be well defined when \(T\) is not sufficiently large relative to \(n\). Our framework is designed to accommodate both conditional heteroskedasticity and unrestricted relative growth of the two panel dimensions.
}

Consider a panel with $n$ individuals and $T$ periods,
\begin{equation*}
 y_{it}=\sum_{j\neq i}\alpha_{ij}y_{jt}+\bm{x}_{it}'\bm{\beta}+\xi_i+\eta_t+\eps_{it},
 \qquad i=1,\ldots,n,\quad t=1,\ldots,T,
\end{equation*}
where $\xi_i$ and $\eta_t$ are individual and time fixed effects, $\bm{x}_{it}\in\mathbb{R}^p$ is a vector of exogenous covariates, and $\alpha_{ij}$ is a  peer-effect coefficient. As in Jung and Liu (2026), the testing problem is
\begin{equation*}
 H_0:\quad \alpha_{ij}=0\quad\text{for all }i\neq j.
\end{equation*}

{ 
The coefficients $\alpha_{ij}$ are allowed to differ across ordered pairs, so interactions need not be symmetric or have a common sign. Following the peer-covariate instrument construction of \citet{jung2026testing}, we treat every other individual as a potential peer and use that individual's exogenous characteristics as candidate instruments. With $p$ covariates, this yields $K=n\left(n-1\right)p$ candidate instruments for $N=nT$ observations. The statistical problem is therefore high-dimensional even when $p$ is fixed.

We propose a leverage-adjusted quadratic-form test. We first remove the two-way fixed effects and estimate the restricted model by ordinary least squares. We then construct a Gram kernel from the residualized peer covariates and adjust its diagonal to account for the dependence introduced by this residualization. The adjustment ensures that the resulting numerator has exactly zero conditional mean under the null, including under conditional heteroskedasticity. The studentized statistic has a standard-normal null limit as $n\wedge T\to\infty$, with no further restriction on the relative growth of $n$ and $T$. Thus short, balanced, and long panels are covered by the same procedure.  

The paper relates to several strands of the literature. One concerns identifying social interactions when network information is available or can be recovered from other observations. \citet{bramoulle2009identification} exploit restrictions implied by an observed network, while \citet{blume2015linear} study identification in linear social-interaction models more generally. With unobserved links, \citet{lewbel2023social} develop identification using a cross section of many small networks, and \citet{depaula2025identifying} give conditions for recovering a time-invariant interaction network from panel data. {These approaches address the task of identifying the network or structural interaction parameters.} We instead avoid estimating the network and formulate an omnibus test using the full set of potential peer relationships.

The closest contribution is \citet{jung2026testing}, who propose an Anderson--Rubin test for the same joint null without specifying the network. Their statistic uses an orthogonal projection onto the instrument space. Its full-network implementation is subject to rank and many-instrument restrictions that lead to a long-panel requirement. In addition, their fixed-effects result relies on homoskedastic idiosyncratic shocks: the within transformation creates dependence that prevents ordinary diagonal deletion from centering a heteroskedastic quadratic form. Our procedure retains their network-unspecified testing objective but replaces the projection kernel by a raw Gram kernel and   corrects the residualization-induced centering error.  

A second strand studies inference with many instruments or many tested restrictions. The many-instrument asymptotics of \citet{bekker1994alternative} provide an alternative to approximations that hold the instrument dimension fixed. Jackknife methods developed by \citet{chao2012asymptotic}, \citet{crudu2021inference}, and \citet{mikusheva2022inference} address heteroskedasticity and, in the latter contributions, identification-robust testing. \citet{anatolyev2023testing} study testing many restrictions under heteroskedasticity, and \citet{mikusheva2024weak} examine how partialling out many exogenous regressors interacts with jackknife centering. {Accommodating more instruments than observations has also been investigated in \citet{dovi2024ridge}, which achieves this through a ridge-regularized jackknifed Anderson--Rubin statistic.} Our contribution is the joint treatment of an unspecified dyadic network, two-way fixed effects, and conditional heteroskedasticity through an unregularized Gram kernel, together with panel conditions that do not impose an ordering on $n$ and $T$.

The quadratic aggregation of peer-covariate moments also relates to high-dimensional mean testing. \citet{chen2010two} use cross-products to estimate a squared mean difference without inverting a high-dimensional covariance matrix. Here the corresponding idea is to aggregate peer-covariate moments through their squared Euclidean norm. A direct transfer of the usual off-diagonal construction is not valid, however, because the residuals are estimated after removing fixed effects and exogenous covariates. This issue is related to the effect of nuisance estimation on heteroskedasticity-robust inference documented by \citet{stock2008heteroskedasticity} and \citet{cattaneo2018inference}. The latter use the inverse of a squared residual-maker matrix to correct variance estimation with many covariates. We use the same underlying residualization identity for a different purpose: choosing a diagonal subtraction that makes the effective test kernel exactly centered in the error coordinates.

Two technical challenges arise. Residualization for fixed effects and observed covariates complicates the centering of quadratic-form statistics under heteroskedasticity, while the large and dependent collection of candidate peer instruments precludes conventional projection-based methods when its dimension exceeds the sample size. We address the first issue through a leverage adjustment that delivers exact conditional centering without estimating observation-specific variances. For the second, we exploit the block structure of the dyadic instruments to establish a weak convergence result and consistent studentization for an quadratic-form statistic. The resulting test accommodates conditional heteroskedasticity and remains valid without imposing a relative-rate restriction on the cross-sectional and time dimensions.

The resulting theory has two further implications. {We also establish consistency against alternatives.} The leading signal is a squared moment associated with the cross-sectionally varying part of the network multiplier; this explains the right-tail rejection rule without requiring all peer coefficients to be positive. Separately, we extend null validity to covariates that are independent across individuals but stationary and weakly dependent over time. Temporal dependence changes the scale of the numerator, but the same feasible variance estimator adapts to that change. This extension requires neither a change in the form of the test nor an additional long-run variance estimator.


The remainder of the paper is organized as follows. Section~\ref{sec:methodology} constructs the test. Section~\ref{sec:theory} presents the null validity and consistency results, and Section~\ref{sec:weakXextension} develops the weak-dependence extension. Section~\ref{sec:simulation} and \ref{sec:application} give  the Monte Carlo experiments and real data analysis.  
Section~\ref{sec:conclusion} concludes with directions for further work. Proofs and technical lemmas follow.
}

\section{Methodology}\label{sec:methodology}
We stack observations in time-major order. For each \(t\), let
\(
\bm y_t=(y_{1t},\ldots,y_{nt})',\) 
\(\bm\varepsilon_t=(\varepsilon_{1t},\ldots,\varepsilon_{nt})',
\) $\mathbf X_t
 :=\left(\bm x_{1t},\ldots,\bm x_{nt}\right)'
 \in\mathbb R^{n\times p}$,
and define
\(
\bm y=(\bm y_1',\ldots,\bm y_T')',\) 
\(
\bm\varepsilon=(\bm\varepsilon_1',\ldots,\bm\varepsilon_T')',
\)
\(
 \mathbf X
 =\left(\mathbf X_1',\ldots,\mathbf X_T'\right)'\in\mathbb R^{N\times p}.
\)
{Under time-major stacking, $\ell=(t-1)n+i$ corresponds to observation $(i,t)$; we write
$\eps_\ell=\eps_{it}$ and $\bm x_\ell=\bm x_{it}$.}
We write
\(
\bm\eta=(\eta_1,\ldots,\eta_T)',\)
\(
\bm\xi=(\xi_1,\ldots,\xi_n)'.
\)
The information set used throughout is
\(
\Fcal:=\sigma(\mathbf X,\bm\eta).
\)
For a matrix, \(\diag(\cdot)\) extracts its diagonal as a vector,
whereas for a vector, \(\Diag(\cdot)\) forms the corresponding
diagonal matrix. We use \(\circ\) to denote the Hadamard
(elementwise) product, and \(\|\cdot\|_F\) and
\(\|\cdot\|_{\mathrm{op}}\) for the Frobenius and operator norms,
respectively.
For a matrix $\mathbf A=(a_{ij})$, define
\(
 \|\mathbf A\|_1:=\max_j\sum_i|a_{ij}|,
 \|\mathbf A\|_\infty:=\max_i\sum_j|a_{ij}|.
\)
For a scalar random variable $Z$,
\(
 \|Z\|_{L^q}:=\left(\mathbb E|Z|^q\right)^{1/q},\) 
and its sub-Gaussian
norm is
\(
\|Z\|_{\psi_2}
:=
\inf\left\{
c>0:
\mathbb E\exp\left(Z^2/c^2\right)\le2
\right\}.
\)
For a random vector \(\bm Z\), we use \(
 \|\bm Z\|_{L^q}
 :=\left\{\mathbb E\left(\|\bm Z\|_2^q\right)\right\}^{1/q},
\) and 
\(
\|\bm Z\|_{\psi_2}
:=
\sup_{\|\bm a\|_2=1}
\|\bm a'\bm Z\|_{\psi_2}.
\)


The construction of the proposed test proceeds in three steps: fitting the model under the no-peer-effects restriction, forming a quadratic aggregate of candidate peer-covariate moments, and correcting its centering before studentization. Throughout, a \emph{restricted residual} is a residual from the model fitted with all $\alpha_{ij}$ set to zero, whereas $\eps_{it}$ denotes the error in the structural equation.  

When the network is completely unknown, every $j\neq i$ is treated as a potential peer of $i$. Candidate instruments are constructed from the exogenous characteristics of potential peers. Let the stacked candidate peer-IV matrix be
\(
 \mathbf{Z}\in\mathbb{R}^{N\times K}\) where $ N=nT$,
and $K$ may grow rapidly with $n$. 
{\color{black}For the completely unknown-network case, index the candidate-IV coordinates by triples $\left(r,j,k\right)$ with $r\ne j$ and $1\le k\le p$.  The row $\bm{z}_{it}'$ of $\mathbf{Z}$ is defined by
\begin{equation}
 z_{it,\left(r,j,k\right)}
 =\mathbf 1\left\{r=i\right\}\,\mathbf 1\left\{j\ne i\right\}\,x_{jt,k}.
 \label{eq:canonicalZ}
\end{equation}
Thus each recipient $i$ has its own dyadic block and uses the contemporaneous exogenous characteristics of every potential peer $j\ne i$. Under this canonical  construction,  $K=n\left(n-1\right)p$, which may greatly exceed $N=nT$.} 
Here, we do \textbf{not} require $K<N$, nor do we require $\mathbf{Z}$ to have full column rank.

Let
\(
 \mathbf{M}_T=\mathbf{I}_T-T^{-1}\bm{1}_T\bm{1}_T'\), \(
 \mathbf{M}_n=\mathbf{I}_n-n^{-1}\bm{1}_n\bm{1}_n'\), 
and define the two-way within transformation
$\mathbf{M}=\mathbf{M}_T\otimes \mathbf{M}_n$.
Let $\mathbf{X}$ be the $N\times p$ stacked matrix of the $\bm{x}_{it}$'s and set $\mathbf{X}^*=\mathbf{M}\mathbf{X}$. Under $H_0$, the within-transformed restricted model is
\begin{equation*}
 \bm{y}^*=\mathbf{X}^*\bm{\beta}+\mathbf{M}\bm{\varepsilon},
 \quad \bm{y}^*=\mathbf{M}\bm{y}.
\end{equation*}
Hence, the  OLS estimator is
\[
 \widehat{\bm{\beta}}=\left(\mathbf{X}^{*'}\mathbf{X}^*\right)^{-1}\mathbf{X}^{*'}\bm{y}^*.
\]
Define $\mathbf{P}_X=\mathbf{X}^*\left(\mathbf{X}^{*'}\mathbf{X}^*\right)^{-1}\mathbf{X}^{*'}$ and $\mathbf{R}=\mathbf{M}-\mathbf{P}_X$. 
Since $\mathbf{M}\mathbf{P}_X=\mathbf{P}_X\mathbf{M}=\mathbf{P}_X$, $\mathbf{R}$ is symmetric and idempotent. The OLS residual can be written as
$\widehat{\bm{u}}=\bm{y}^*-\mathbf{X}^*\widehat{\bm{\beta}}=\mathbf{R}\bm{y}$. 
Under $H_0$, $\mathbf{R}$ annihilates both the two-way fixed effects and $\mathbf{X}\bm{\beta}$, yielding the following key exact identity
\begin{equation}
\widehat{\bm{u}}=\mathbf{R}\bm{\varepsilon}.
 \label{eq:uhat_exact}
\end{equation}
Thus, there is no need to treat $\widehat{\bm{\beta}}-\bm{\beta}$ as a separate remainder. The OLS estimation effect is absorbed exactly into the residual-maker $\mathbf{R}$,
{and the price is that the entries of $\widehat{\bm u}$ are dependent.}

\subsection{The naive U-statistic}

Since $\mathbf R\mathbf X^*=\mathbf0$, the exogenous covariates are annihilated by residualization and need not be included in the instrument dictionary. In fact, residualizing $[\mathbf X^*,\mathbf Z]$ gives $[\mathbf0,\mathbf R\mathbf Z]$, with exactly the same Gram matrix as the residualized peer instruments alone. We therefore set
$\mathbf{Q}=\mathbf{R}\mathbf{Z}\in\mathbb{R}^{N\times K}$.
 Since $\mathbf{R}^2=\mathbf{R}$,
 $\mathbf{R}\mathbf{Q}=\mathbf{Q}$.
Let $\bm{q}_\ell'$ denote the $\ell$th row of $\mathbf{Q}$, where $\ell=1,\ldots,N$ indexes the stacked observation $\left(i,t\right)$.

The use of $\mathbf Q\mathbf Q'$ corresponds to summing the squared sample moments between the instruments and the restricted residuals. This is the same basic reason that cross-product statistics can remain defined when a high-dimensional covariance matrix cannot be inverted \citep{chen2010two}.
If $\bm{\varepsilon}$ were directly observed and there were no fixed effects or nuisance estimation, a natural cross-product statistic would be
\begin{equation*}
 U_N^{\mathrm{oracle}}
 :=\sum_{\ell\neq \ell'}\eps_{\ell}\eps_{\ell'} \bm{q}_{\ell}'\bm{q}_{\ell'}
 =\bm{\varepsilon}'\left\{\mathbf{Q}\mathbf{Q}'-\Diag\left(\diag\left(\mathbf{Q}\mathbf{Q}'\right)\right)\right\}\bm{\varepsilon}.
\end{equation*}

This construction uses identity weighting rather than inverse-Gram weighting and removes the diagonal terms. 
The instrument residualization and leverage correction below distinguish our statistic from the projection-based construction of \citet{jung2026testing}.
The purpose here is to avoid  controlling of the ratio $K/N$ and 
requiring the residualized instrument matrix $\mathbf{Q}$ to have full column rank. 

If $\mathbb{E}\left(\eps_{\ell}\mid\Fcal\right)=0$ and the $\eps_{\ell}$'s are conditionally independent, $U_N^{\mathrm{oracle}}$ has zero conditional mean under $H_0$.
In practice, however, only $\widehat{\bm{u}}=\mathbf{R}\bm{\varepsilon}$ is available. The naive plug-in residual statistic is
\begin{equation}
 U_N^{\mathrm{naive}}
 :=\sum_{\ell\neq \ell'}\widehat{\bm{u}}_{\ell}\widehat{\bm{u}}_{\ell'} \bm{q}_{\ell}'\bm{q}_{\ell'}=\widehat{\bm{u}}'\left\{\mathbf{Q}\mathbf{Q}'-\Diag\left(\diag\left(\mathbf{Q}\mathbf{Q}'\right)\right)\right\}\widehat{\bm{u}}.
 \label{eq:naive}
\end{equation}
There is a crucial problem. Although the matrix between the two residual vectors in \eqref{eq:naive} has zero diagonal, under $H_0$,
\[
 U_N^{\mathrm{naive}}
 =\bm{\varepsilon}'\mathbf{R}\left\{\mathbf{Q}\mathbf{Q}'-\Diag\left(\diag\left(\mathbf{Q}\mathbf{Q}'\right)\right)\right\}\mathbf{R}\bm{\varepsilon}.
\]
 Since $\mathbf{R}\mathbf{Q}=\mathbf{Q}$,
\begin{equation*}
 \diag\left[\mathbf{R}\left\{\mathbf{Q}\mathbf{Q}'-\Diag\left(\diag\left(\mathbf{Q}\mathbf{Q}'\right)\right)\right\}\mathbf{R}\right]
 =\diag\left(\mathbf{Q}\mathbf{Q}'\right)-\left(\mathbf{R}\circ \mathbf{R}\right)\diag\left(\mathbf{Q}\mathbf{Q}'\right).
\end{equation*}
Let $\Fcal:=\sigma\left(\mathbf X,\bm\eta\right)$ and $\sigma_{\ell}^2=\mathbb{E}\left(\eps_{\ell}^2\mid\Fcal\right)$. Under conditional heteroskedasticity with $\mathbf{\Omega}=\Diag\left(\sigma_{1}^2,\ldots,\sigma_{N}^2\right)$, 
\[
 \mathbb{E}\left(U_N^{\mathrm{naive}}\mid\Fcal\right)
 =\sum_{\ell=1}^N
 \left\{\left[\diag\left(\mathbf{Q}\mathbf{Q}'\right)\right]_{\ell}
 -\left[\left(\mathbf{R}\circ \mathbf{R}\right)\diag\left(\mathbf{Q}\mathbf{Q}'\right)\right]_{\ell}\right\}\sigma_{\ell}^2,
\]
which is generally nonzero. Hence, {deleting only the diagonal of $\mathbf{Q}\mathbf{Q}'$ is not enough to obtain exact centering after two-way FE and OLS residualization}.

{The failure arises because diagonal deletion and residualization do not commute. A correction performed only in residual coordinates can leave nonzero diagonal entries in error coordinates, where heteroskedasticity determines the conditional mean. This is why the usual jackknife adjustment must be modified after two-way fixed-effects removal.}

\subsection{Leverage-adjusted diagonal correction}\label{subsec:Leverage-adjusted}
The correction is chosen so that, in the coordinates of the independent shocks $\bm{\varepsilon}$, the effective quadratic-form matrix has exactly zero diagonal. Let $\bm{\tau}\in\mathbb{R}^N$ solve
\begin{equation}
 \left(\mathbf{R}\circ \mathbf{R}\right)\bm{\tau}=\diag\left(\mathbf{Q}\mathbf{Q}'\right).
 \label{eq:tauEquation}
\end{equation}
The existence and stability of this solution are guaranteed by Lemma 2. 
The matrix $\mathbf R\circ\mathbf R$ describes how diagonal contributions are redistributed by residualization. Solving \eqref{eq:tauEquation} reverses this redistribution for the diagonal of $\mathbf Q\mathbf Q'$. Related squared-residual-maker corrections appear in heteroskedastic variance estimation with many covariates \citep{cattaneo2018inference}; preserving jackknife centering after partialling out controls is also central to \citet{mikusheva2024weak}.

The proposed numerator is
\begin{equation}
 S_N
 =\widehat{\bm{u}}'\left\{\mathbf{Q}\mathbf{Q}'-\Diag\left(\bm{\tau}\right)\right\}\widehat{\bm{u}},
 \qquad
 \bm{\tau}=\left(\mathbf{R}\circ \mathbf{R}\right)^{-1}\diag\left(\mathbf{Q}\mathbf{Q}'\right).
 \label{eq:SN}
\end{equation}
Under $H_0$, using \eqref{eq:uhat_exact},
$S_N
 =\bm{\varepsilon}'\mathbf{R}\left\{\mathbf{Q}\mathbf{Q}'-\Diag\left(\bm{\tau}\right)\right\}\mathbf{R}\bm{\varepsilon}.$
By \eqref{eq:tauEquation},
\begin{equation*}
 \diag\left[\mathbf{R}\left\{\mathbf{Q}\mathbf{Q}'-\Diag\left(\bm{\tau}\right)\right\}\mathbf{R}\right]
 =\diag\left(\mathbf{Q}\mathbf{Q}'\right)-\left(\mathbf{R}\circ \mathbf{R}\right)\bm{\tau}
 =0.
\end{equation*}
Therefore, if $b_{\ell\ell'}$ denotes the $\left(\ell,\ell'\right)$th element of the effective matrix
$\mathbf B_N=\mathbf{R}\left\{\mathbf{Q}\mathbf{Q}'-\Diag\left(\bm{\tau}\right)\right\}\mathbf{R}$, then
\begin{equation*}
 S_N=\sum_{\ell\neq \ell'}b_{\ell\ell'}\eps_{\ell}\eps_{\ell'},
\end{equation*}
which is a zero-diagonal quadratic form in the conditionally independent errors. 
Note that if there are no fixed effects and no nuisance estimation, so that $\mathbf{R}=\mathbf{I}_N$, then \eqref{eq:tauEquation} gives $\bm{\tau}=\diag\left(\mathbf{Q}\mathbf{Q}'\right)$, and \eqref{eq:SN} reduces exactly to
\(
 S_N=\sum_{\ell\neq \ell'}\eps_{\ell}\eps_{\ell'}\bm{q}_{\ell}'\bm{q}_{\ell'}.
\)

\paragraph{Variance and studentization.}
  Since $b_{\ell\ell}=0$, conditional independence gives
\begin{equation*}
 V_N:=\Var\left(S_N\mid\Fcal\right)
 =2\sum_{\ell\neq \ell'}b_{\ell\ell'}^2\sigma_{\ell}^2\sigma_{\ell'}^2.
\end{equation*}
A natural feasible estimator is
\begin{equation*}
 \widehat V_N
 =2\sum_{\ell\neq \ell'}b_{\ell\ell'}^2\widehat{\bm{u}}_{\ell}^2\widehat{\bm{u}}_{\ell'}^2.
\end{equation*}
Therefore, the final test statistic is
$ T_N= {S_N}/{{\widehat V_N^{1/2}}}.$
Our construction uses only $\mathbf{Q}\mathbf{Q}'$ and never requires $\mathbf{Q}'\mathbf{Q}$ to be invertible, so $K$ may be much larger than $N=nT$. This implies that our test statistic remains   well defined whether $T<n$, $T\asymp n$, or $T>n$.

{
In addition, note that both $\mathbf B_N$ and $\widehat V_N$ are obtained from the observed covariates and restricted residuals. Hence, implementation does not require estimating the  conditional variance of $\varepsilon_{it}$  or the peer-effect coefficients.  The null approximation below yields a test that rejects for $T_N>z_{1-\alpha}$, where $z_{1-\alpha}$ is the standard-normal quantile, and the corresponding theoretical justification is given in Section~\ref{sec:theory}.
}

\section{Theory}\label{sec:theory}
{
We first establish null validity and then prove consistency under  alternatives. 
}

\subsection{Under the null}

\begingroup
\begin{assumption}\label{ass:error}
Conditional on $\Fcal$, the idiosyncratic errors
$\{\eps_{it}:1\le i\le n,\ 1\le t\le T\}$ are mutually independent and satisfy
\[
 \mathbb E\left(\eps_{it}\mid\Fcal\right)=0,
 \qquad
 \mathbb E\left(\eps_{it}^2\mid\Fcal\right)
 =\sigma^2\left(\bm x_{it}\right),
 \qquad
 \max_{i,t}\mathbb E\left(\eps_{it}^8\mid\Fcal\right)
 \le C_\eps<\infty.
\]
For a constant $C_\sigma<\infty$,
\(
 0\le\sigma^2\left(\bm x\right)\le C_\sigma\), and
\(\bar\sigma^2:=\mathbb E\left\{\sigma^2\left(\bm x_{it}\right)\right\}>0.
\)
\end{assumption}
\endgroup
{
{Assumption~\ref{ass:error} permits nonlinear heteroskedasticity in the covariates and allows $\sigma^2\left(\bm x_{it}\right)=0$ with positive probability, while conditional independence of $\eps_{it}$ across observations remains a restriction.} {Section~\ref{sec:weakXextension} further relaxes temporal independence of the covariates.} The upper bound and eighth-moment requirement control products of residuals in the quadratic form and its variance estimator. {Finite conditional moments are familiar in many-instrument inference \citep{chao2012asymptotic,mikusheva2022inference}.} 
}

\begin{assumption}\label{ass:xprimitive}
The covariate dimension $p$ is fixed and $ n\wedge T\to\infty$. The covariates $\bm{x}_{it}\in\mathbb{R}^{p}$ are i.i.d.\ over $\left(i,t\right)$ and satisfy $\mathbb{E}\left(\bm{x}_{it}\right)=0$, $\mathbf{\Sigma}_x=\mathbb{E}\left(\bm{x}_{it}\bm{x}_{it}'\right)$, $ \sup_{i,t}\mathbb{E}\left\{\|\bm{x}_{it}\|_2^{q_x}\right\}\le C_{x,q}<\infty$ for some $q_x:=8+\nu_x$, $\nu_x>0$.
The covariance matrix satisfies that
\begin{equation}
 0<c_x\le \lambda_{\min}\left(\mathbf{\Sigma}_x\right)
 \le \lambda_{\max}\left(\mathbf{\Sigma}_x\right)\le C_x<\infty.
 \label{eq:Sigmaxbounds}
\end{equation}
\end{assumption}
{
The eigenvalue bounds in \eqref{eq:Sigmaxbounds} ensure that the fixed-dimensional linear regression is nondegenerate. {Such design regularity plays the same role as the nonsingularity conditions in the many-instrument analyses of \citet{crudu2021inference} and \citet{jung2026testing}.}    Temporal dependent $\bm x_{it}$ is also considered in Section~\ref{sec:weakXextension}.
}

\begin{theorem}\label{thm:null}
Suppose $H_0$ holds, Assumptions \ref{ass:error} and \ref{ass:xprimitive} are satisfied, and $\mathbf{Z}$ is the canonical choice   in \eqref{eq:canonicalZ}.
Then,
\begin{equation*}
{T_N\Rightarrow \mathcal{N}\left(0,1\right).}
\end{equation*}
\end{theorem}
{
Exact centering and asymptotic normality have different roles in this result. The first follows from the zero diagonal of $\mathbf B_N$ in error coordinates. The second requires that no small part of the kernel dominate the quadratic form, together with consistent studentization. The proofs verify these properties using the dyadic block structure and vanishing nuisance leverage, followed by a generalized quadratic-form central limit theorem \citep{dejong1987central}.
}

Theorem \ref{thm:null} imposes no rank restriction of the form $K<N$. Under Assumption \ref{ass:xprimitive},  the only panel-growth requirement is $n\wedge T\to\infty$. Hence the theorem covers short, balanced, and long panels, including $T/n\to0$, $T/n\to c\in\left(0,\infty\right)$, and $T/n\to\infty$, without any restriction on the relative rates of $n$ and $T$.

\begingroup 
\begin{remark}\label{rem:divergingp}
A diverging-dimensional extension is available by replacing the fixed-$p$ and vector-moment requirements in Assumption~\ref{ass:xprimitive} with
\(
 p=p_N\), \(p/T\to0\), and \(
 \sup_{\|\bm a\|_2=1}\|\bm a'\bm x_{it}\|_{\psi_2}\le K_x<\infty.
\)
The covariates remain i.i.d. over $(i,t)$, centered, and satisfy the uniform eigenvalue bounds \eqref{eq:Sigmaxbounds}. 
Under these conditions the conclusion of Theorem~\ref{thm:null} remains valid; a proof is given in Lemma~20. 
\end{remark}
\endgroup


\subsection{Under the alternative}

Since we   fit the model with peer coefficients restricted to zero, the resulting residuals   contain both the error component and the part of the peer interaction omitted from that regression under the alternative. 

Let $\mathbf{A}_n=[\alpha_{ij}]_{i,j=1}^n$ denote the time-invariant peer-effect matrix, with zero diagonal, and define the $N\times N$ panel network operator
$\mathbf{A}_N=\mathbf{I}_T\otimes \mathbf{A}_n.$
Let $\bm{f}_N=\bm{1}_T\otimes\bm{\xi}+\bm{\eta}\otimes\bm{1}_n$ and $\bm{m}_N=\mathbf{X}\bm{\beta}+\bm{f}_N.$
{A sequence of alternatives can be written as}
\begin{equation}\label{eq:outcome}
 H_{1,N}: \bm{y}={\mathbf{A}_N\bm y}+\bm{m}_N+\bm{\varepsilon}.
\end{equation}
Assume $\mathbf I_N-\mathbf A_N$ is nonsingular and define
\[
 \mathbf H_N=\left(\mathbf I_N-\mathbf A_N\right)^{-1},
 \qquad
 \mathbf\Gamma_N=\mathbf H_N-\mathbf I_N
 =\mathbf A_N\left(\mathbf I_N-\mathbf A_N\right)^{-1}.
\] 
Then
$\bm{y}=\mathbf{H}_N\left(\bm{m}_N+\bm{\varepsilon}\right).$
Because $\mathbf{R}\bm{m}_N=0$, the restricted residual under $H_{1,N}$ has the exact decomposition
\begin{equation*}
{
 \widehat{\bm{u}}=\mathbf{R}\bm{y}
 =\mathbf{R}\bm{\varepsilon}+\bm{\delta}_N+\mathbf{D}_N\bm{\varepsilon},
 }
\end{equation*}
where
\begin{equation}
 \bm{\delta}_N=\mathbf{R}\mathbf{\Gamma}_N\bm{m}_N,
 \quad
 \mathbf{D}_N=\mathbf{R}\mathbf{\Gamma}_N,
 \quad
 \mathbf{L}_N=\mathbf{R}+\mathbf{D}_N=\mathbf{R}\mathbf{H}_N.
 \label{eq:deltaDL}
\end{equation}

Recall the null effective kernel $\mathbf{B}_N$ defined in Section \ref{subsec:Leverage-adjusted}. Define
\begin{equation}
 {\mathbf{K}_N=\mathbf{H}_N'\mathbf{B}_N\mathbf{H}_N},
 \quad
 \bm{a}_N=\mathbf{H}_N'\mathbf{B}_N\bm{\delta}_N,
 \quad
 \mathbf{E}_N=\mathbf{K}_N-\mathbf{B}_N.
 \label{eq:KaE}
\end{equation}
Then, the numerator admits equivalent exact decompositions:
\begin{align}
 S_N
 &=\bm{\delta}_N'\mathbf{B}_N\bm{\delta}_N+2\bm{a}_N'\bm{\varepsilon}+\bm{\varepsilon}'\mathbf{K}_N\bm{\varepsilon}
 \label{eq:altexact1}=S_{0N}+\bm{\delta}_N'\mathbf{B}_N\bm{\delta}_N+2\bm{a}_N'\bm{\varepsilon}+\bm{\varepsilon}'\mathbf{E}_N\bm{\varepsilon},
\end{align}
where $S_{0N}=\bm{\varepsilon}'\mathbf{B}_N\bm{\varepsilon}$.
Under $H_0$, $\mathbf{H}_N=\mathbf{I}_N$, $\mathbf{\Gamma}_N=0$, $\bm{\delta}_N=0$, $\mathbf{L}_N=\mathbf{R}$, $\mathbf{K}_N=\mathbf{B}_N$, and $\mathbf{E}_N=0$, so \eqref{eq:altexact1} reduces to the null representation in Section \ref{subsec:Leverage-adjusted}. The following exact decomposition distinguishes the conditional mean shift from the change in covariance induced by network interactions.

\begin{proposition}\label{prop:altmom}
Suppose Assumption \ref{ass:error}  holds. Let $\mu_{3\ell}=\mathbb{E}\left(\eps_{\ell}^3\mid\Fcal\right)$ and $\kappa_{4\ell}=\mathbb{E}\left(\eps_{\ell}^4\mid\Fcal\right)-3\sigma_{\ell}^4$,
and let $k_{\ell\ell}$ be the $\ell$th diagonal element of $\mathbf{K}_N$. Then, {under the alternatives $H_{1,N}$},
\begin{align}
 \label{eq:altmean}
 &\mathfrak m_N:=\mathbb{E}\left(S_N\mid\Fcal\right)
 =\bm{\delta}_N'\mathbf{B}_N\bm{\delta}_N+\tr\left(\mathbf{E}_N\mathbf{\Omega}\right)=\bm{\delta}_N'\mathbf{B}_N\bm{\delta}_N+\tr\left(\mathbf{K}_N\mathbf{\Omega}\right).
 \\
 &\mathfrak v_N
 :=\Var\left(S_N\mid\Fcal\right)
 =4\bm{a}_N'\mathbf{\Omega} \bm{a}_N+2\tr\left(\mathbf{K}_N\mathbf{\Omega} \mathbf{K}_N\mathbf{\Omega}\right) +\sum_{\ell=1}^N\kappa_{4\ell}k_{\ell\ell}^2
 +4\sum_{\ell=1}^N\mu_{3\ell}\left(\bm{a}_N\right)_{\ell} k_{\ell\ell}.
 \label{eq:altvar}
\end{align}
\end{proposition}

The conditional mean in \eqref{eq:altmean} admits the decomposition
\begin{equation*}
 \mathfrak m_N
 =
 \underbrace{
 \bm{\delta}_N'\mathbf{B}_N\bm{\delta}_N
 }_{\text{mean-shift component}}
 +
 \underbrace{
 \tr\left(\mathbf{E}_N\mathbf{\Omega}\right)
 }_{\text{covariance-feedback component}}.
\end{equation*}
The first term captures the systematic deviation from the null model induced by the propagation of covariates and fixed effects through the peer network. The second term arises because, under the alternative, the peer network transmits the shocks $\varepsilon_{it}$ across individuals. As a result, the stochastic component of the restricted residuals $\widehat{\bm u}$ has conditional covariance
\(
\mathbf R\mathbf H_N\mathbf\Omega\mathbf H_N'\mathbf R,
\)
rather than the null covariance $\mathbf R\mathbf\Omega\mathbf R$. This term vanishes identically under $H_0$, for which $\mathbf H_N=\mathbf I_N$, but can be nonzero under the alternative.

\begingroup 
\begin{assumption}\label{ass:altstable}
The peer-effect matrix $\mathbf A_n=\left[\alpha_{ij}\right]_{i,j=1}^n$ is deterministic and satisfies $\left(\mathbf A_n\right)_{ii}=0$. There exist a deterministic sequence $\rho_N>0$ and a constant $\kappa\in\left(0,1\right)$ such that
\begin{equation}
 \max_i\sum_{j=1}^n\left|\left(\mathbf A_n\right)_{ij}\right|
 \vee
 \max_j\sum_{i=1}^n\left|\left(\mathbf A_n\right)_{ij}\right|
 \le \rho_N\le1-\kappa.
 \label{eq:networkrowscols}
\end{equation}
The slope vector is deterministic and satisfies
$\|\bm\beta\|_2\le C_\beta$.
\end{assumption}
Assumption~\ref{ass:altstable} controls the total absolute peer effect entering and leaving each individual. In particular,
$\|\mathbf A_n\|_{\infty}\vee\|\mathbf A_n\|_1\le\rho_N$, thus
$\|\mathbf A_n\|_{\mathrm{op}}\le\rho_N\le1-\kappa$. 
Therefore, \(\mathbf I_n-\mathbf A_n\) is nonsingular and its inverse admits a uniformly bounded Neumann-series representation, thus the outcome equation \eqref{eq:outcome} admits a unique reduced-form solution.  The same condition also yields an order-$\rho_N$ bound for $\mathbf\Gamma_n=\left(\mathbf I_n-\mathbf A_n\right)^{-1}-\mathbf I_n$. Bounded absolute row and column sums are commonly used to control spatial multipliers; see, for example, \citet{lee2004asymptotic}.
\endgroup

\begingroup 
\begin{assumption}\label{ass:timeeffect}
The time-effect process is mean-zero, strictly stationary, and independent of the entire covariate array. For some $q_\eta>8$,
\(
 \sup_t\mathbb E|\eta_t|^{q_\eta}\le C_\eta<\infty.
\)
In addition, one of the following two conditions holds:
\begin{enumerate}[label=(\roman*),leftmargin=2.4em]
\item The strong-mixing coefficients of the time-effect process obey
\begin{equation}
 \sup_N\sum_{h=1}^{\infty}(h+1)
 \alpha_{\eta,N}(h)^{1-8/q_\eta}<\infty.
 \label{eq:mixingLocal}
\end{equation}
\item The time effect has a causal representation in i.i.d. innovations, independent of the covariate array. If $\eta_t^{*(h)}$ is obtained by replacing the innovation at $t-h$ by an independent copy and
$\delta_{\eta,q_\eta,N}(h):=\sup_t\|\eta_t-\eta_t^{*(h)}\|_{q_\eta}$, then
\begin{equation}
 \sup_N\sum_{h=0}^{\infty}(h+1)
 \delta_{\eta,q_\eta,N}(h)<\infty.
 \label{eq:physicalLocal}
\end{equation}
\end{enumerate}
\end{assumption}

\endgroup


We next give sufficient conditions for consistency against  alternatives.  

\begingroup 
\begin{assumption}\label{ass:separated}
For a constant $c_{\beta,*}>0$, $\|\mathbf\Sigma_x\bm\beta\|_2\ge c_{\beta,*}$. The network matrix satisfies
\begin{equation}
 \frac{\|\bm1_n'\mathbf A_n\|_2^2}
 {n\|\mathbf A_n\|_F^2}\longrightarrow0
 \label{eq:primitiveNetworkSpread}
\end{equation}
and
\begin{equation}
 \frac{\sqrt{nT}(n+T)
       +T\sqrt n\,\|\mathbf M_n\mathbf A_n\bm1_n\|_2^2}
 {T^2\|\mathbf A_n\|_F^2}
 \longrightarrow0.
 \label{eq:primitiveSeparation}
\end{equation}
\end{assumption}
Condition~\eqref{eq:primitiveNetworkSpread} controls the component of the network that is common across recipients: its numerator divided by $n$ is the squared Frobenius norm of $n^{-1}\bm1_n\bm1_n'\mathbf A_n$. Together with stability and the zero diagonal of $\mathbf A_n$, it ensures that a nonvanishing fraction of network energy remains in the peer-covariate moments after nuisance regression; see Lemma~11. 
Condition~\eqref{eq:primitiveSeparation} requires this signal to dominate the quadratic-form fluctuations and the additional variation from heterogeneous propagation of the common time effect. The latter is controlled directly by the dispersion of the row sums, $\|\mathbf M_n\mathbf A_n\bm1_n\|_2^2$, and is zero for a constant-row-sum network. 
\endgroup

\begingroup 
\begin{theorem}\label{thm:consistency}
{Under Assumptions \ref{ass:error}, \ref{ass:xprimitive}, \ref{ass:altstable}, \ref{ass:timeeffect}, and \ref{ass:separated},}
\(
 T_N\xrightarrow{p}+\infty.
\)
\end{theorem}
\endgroup

 Hence, the rejection probability of the right-tail test tends to one at every fixed asymptotic level. 

\begingroup 
\begin{remark}[Examples of  alternatives]\label{rem:separatedNetworks}
The following stable versions of the network configurations discussed by \citet{jung2026testing} illustrate Assumption~\ref{ass:separated}. Throughout, $\|\mathbf\Sigma_x\bm\beta\|_2\asymp1$, $a\ne0$ is fixed with $|a|<1-\kappa$, and $n\wedge T\to\infty$. The restrictions below verify the sufficient conditions of Theorem~\ref{thm:consistency}.
\begin{itemize}[leftmargin=1.6em]
\item[(a)] \emph{A fixed link.}
Let $\mathbf A_n=a\bm e_1\bm e_2'$. Its maximum absolute row and column sums equal $|a|$, and
\[
 \|\mathbf A_n\|_F^2=a^2,\qquad
 \frac{\|\bm1_n'\mathbf A_n\|_2^2}{n\|\mathbf A_n\|_F^2}=\frac1n,
 \qquad
 \|\mathbf M_n\mathbf A_n\bm1_n\|_2^2=a^2(1-n^{-1}).
\]
Thus Assumption~\ref{ass:separated} holds when $T/n\to\infty$.  

\item[(b)] \emph{Bounded degree.}
Let $\mathbf S_n$ be the cyclic shift matrix, with $(\mathbf S_n)_{i,i+1}=1$ for $i<n$ and $(\mathbf S_n)_{n,1}=1$. For a fixed integer $d\ge1$ and $n>d$, set
\(
 \mathbf A_n=a\sum_{k=1}^d\mathbf S_n^k/d.
\)
Every row and column has $d$ links and total absolute weight $|a|$. Here
\[
 \|\mathbf A_n\|_F^2=\frac{na^2}{d},\qquad
 \frac{\|\bm1_n'\mathbf A_n\|_2^2}{n\|\mathbf A_n\|_F^2}=\frac dn,
 \qquad
 \mathbf M_n\mathbf A_n\bm1_n=\bm0.
\]
Condition~\eqref{eq:primitiveSeparation} is equivalent to $n/T^3\to0$.  The same conclusion holds for the two-neighbor circular network $\mathbf A_n=a(\mathbf S_n+\mathbf S_n')/2$.


\item[(c)] \emph{Fully dense, heterogeneous effects.}
For even $n$, let $\bm s_n\in\{-1,1\}^n$ satisfy $\bm1_n'\bm s_n=0$, and define
\(
 \mathbf A_n=a(\bm s_n\bm s_n'-\mathbf I_n)/n.
\)
All off-diagonal entries are nonzero, and the absolute row and column sums are $|a|(n-1)/n$. Moreover,
\[
 \|\mathbf A_n\|_F^2=\frac{a^2(n-1)}n,\qquad
 \frac{\|\bm1_n'\mathbf A_n\|_2^2}{n\|\mathbf A_n\|_F^2}
 =\frac1{n(n-1)},\qquad
 \mathbf M_n\mathbf A_n\bm1_n=\bm0.
\]
Consequently, Assumption~\ref{ass:separated} holds when $T/n\to\infty$.
The arrangement of heterogeneity matters. For the alternative dense matrix
$\mathbf A_n=a\Diag(\bm s_n)(\bm1_n\bm1_n'-\mathbf I_n)/(n-1)$, 
$\|\mathbf M_n\mathbf A_n\bm1_n\|_2^2=na^2$ and
$\|\mathbf A_n\|_F^2=na^2/(n-1)$. In that case \eqref{eq:primitiveSeparation} requires $T/n^{3/2}\to\infty$, reflecting the heterogeneous transmission of the common time effect.

\item[(d)] \emph{Complete homogeneous effects: a degenerate case.}
Let
\(
 \mathbf A_n= {a} (\bm1_n\bm1_n'-\mathbf I_n)/(n-1).
\)
Then, Assumption~\ref{ass:altstable} holds, and $\|\mathbf A_n\|_F^2=na^2/(n-1)$, so \eqref{eq:primitiveSeparation} can hold when $T/n\to\infty$. However,
\[
 \frac{\|\bm1_n'\mathbf A_n\|_2^2}{n\|\mathbf A_n\|_F^2}
 =\frac{n-1}{n}\longrightarrow1,
 \qquad
 \mathbf M_n\mathbf\Gamma_n=-\frac{a}{n-1+a}\mathbf M_n.
\]
Thus \eqref{eq:primitiveNetworkSpread} fails. After removing the fixed effects, the network only rescales the own-covariate regression and the errors. In fact,
\(
 \widehat{\bm u}= {(n-1)} \mathbf R\bm\varepsilon/(n-1+a).
\)
The common scale cancels from $T_N$, so the statistic is identical to its no-peer-effect counterpart for the same covariates and  errors, which is the homogeneous   degeneracy discussed by \citet{jung2026testing}. This example shows why conditions based only on $\|\mathbf A_n\|_F^2$ cannot establish consistency of the test.
\end{itemize}
\end{remark}
\endgroup

\section{Extension to weakly dependent covariates}\label{sec:weakXextension}

The preceding results impose temporal independence on the covariates $\bm x_{it}$.  We now show that the null theory continues to hold when each individual's covariate process is serially dependent.  Throughout this section, Assumption \ref{ass:error} is retained, while Assumption \ref{ass:xprimitive} is replaced by the following condition. 
{
{The relaxation concerns dependence within a covariate trajectory, and entire trajectories remain independent across individuals.} The  conditions below are used to control the random Gram kernel generated by these trajectories.
}

\begingroup 
\begin{assumption}[Weakly dependent covariates]\label{ass:xweak}
The covariate dimension $p$ is fixed and $n\wedge T\to\infty$.  For every $i$, $\{\bm x_{it}\}_{t\in\mathbb Z}$ is a strictly stationary $\mathbb R^p$-valued process, and the individual trajectories $\{\bm x_{i,\cdot}\}_{i=1}^n$ are independent and identically distributed across $i$.   Moreover,
\(
 \mathbb E\left(\bm x_{it}\right)=\bm0\), \(
 \mathbf\Sigma_x:=\mathbb E\left(\bm x_{it}\bm x_{it}'\right)\), 
and \eqref{eq:Sigmaxbounds} holds.  For some $q:=16+\nu$, $\nu>0$,
\(
 \sup_{i,t}\mathbb E\left\{\|\bm x_{it}\|_2^{q}\right\}\le C_{q}<\infty.
\)
There exists a constant $c_\sigma>0$ such that
$
 \inf_{h\ge1}\mathbb E\left[\sigma^2\left(\bm x_{i0}\right)\sigma^2\left(\bm x_{ih}\right)\right]\ge c_\sigma$.
 
In addition, one of the following two weak-dependence conditions is assumed.
\begin{enumerate}[label=(\roman*),leftmargin=2.4em]
 \item \textbf{Strong-mixing.} Let $\alpha_x(h)$ denote the strong-mixing coefficient of an individual trajectory.  We assume
 \begin{equation}
  \sum_{h=1}^{\infty}(h+1)^2\alpha_x(h)^{1-16/q}<\infty.
  \label{eq:weakXmixing}
 \end{equation}
 \item \textbf{Physical-dependence.} For every $i$, the covariate process admits a causal representation
\(
  \bm x_{it}=G_x\left(\ldots,\bm\zeta_{i,t-1},\bm\zeta_{it}\right)\),
 where the innovation sequences $\{\bm\zeta_{it}:t\in\mathbb Z\}$ are i.i.d. over $t$ and independent across $i$.  Let $\bm x_{it}^{*(h)}$ denote the coupled version obtained by replacing $\bm\zeta_{i,t-h}$ with an independent copy.  Define the augmented vectors
 \(
  \bm v_{it}:=\left(\bm x_{it}',\sigma^2\left(\bm x_{it}\right)\right)'\),
  \(
  \bm v_{it}^{*(h)}:=\left(\bm x_{it}^{*(h)'},\sigma^2\left(\bm x_{it}^{*(h)}\right)\right)',
 \)
 and the physical-dependence coefficient
\(
  \delta_{v,q}(h)
  :=\sup_{i,t}\left\|\bm v_{it}-\bm v_{it}^{*(h)}\right\|_{q}\). 
 We assume
 \begin{equation}
  \sum_{h=0}^{\infty}(h+1)^3\delta_{v,q}(h)<\infty.
  \label{eq:weakXphysical}
 \end{equation}
\end{enumerate}
\end{assumption}
\endgroup

{ 
{Both branches of Assumption~\ref{ass:xweak} are alternative sufficient frameworks, not equivalent assumptions.}  
}
\begingroup 
The variance coordinate is included in $\bm v_{it}$ to control temporal dependence of the variance weights without assuming smoothness of $\sigma^2(\cdot)$. {The lag-overlap lower bound is an additional nondegeneracy requirement, since positive average variance alone does not rule out zero overlap at a particular lag.}   
\endgroup

\begin{theorem}\label{thm:nullweak}
Suppose $H_0$ holds, Assumptions \ref{ass:error} and \ref{ass:xweak} are satisfied, and $\mathbf Z$ is the canonical choice in \eqref{eq:canonicalZ}.  Then
\[
 T_N\Rightarrow\mathcal N(0,1).
\]
\end{theorem}
{Theorem \ref{thm:nullweak} requires only $n\wedge T\to\infty$ and imposes no restriction on the relative growth of $n$ and $T$.
Conditional on $\Fcal$, the covariate is fixed and the  shocks remain independent. Consequently, temporal dependence in the covariates changes the coefficients of the quadratic form, not the conditional covariance formula for independent  shocks. This is why the same variance estimator remains applicable, and no long-run variance needs to be estimated.
}

\begin{remark}[Effect of temporal dependence]
\label{rem:weakXinterpretation}
Although Theorem~\ref{thm:nullweak} has the same standard-normal
limit as Theorem~\ref{thm:null}, temporal dependence in the covariates
changes the conditional variance of the numerator $S_N$. 
In fact, 
\begin{align*}
    &\text{Var}(S_N|\mathcal F_N)\\=&\{1+\o_p(1)\}4n(n-1)\sum_{h=1}^{T-1}(T-h)\mathbb E\left\{
\sigma^2(\bm x_{i0})
\sigma^2(\bm x_{ih})
\right\}\left[ {\mathbb E\left\{\left(\bm x_{i0}'\bm x_{ih}\right)^2\right\}}
 +(n-2) \{\mathbb E(\bm x_{i0}'\bm x_{ih})\}^2\right]\\=&2n(n-1)T(T-1)
(\bar\sigma^2)^2
\operatorname{tr}(\mathbf\Sigma_x^2)
+
4n(n-1)(n-2)T\sum_{h=1}^{\infty}\E\left\{
\sigma^2(\bm x_{i0})
\sigma^2(\bm x_{ih})
\right\}\{\mathbb E(\bm x_{i0}'\bm x_{ih})\}^2\\&+\o_p\left(n^2T^2+n^3T\right).
\end{align*}
  The term \({\mathbb E\left\{\left(\bm x_{i0}'\bm x_{ih}\right)^2\right\}}\) is the second moment of a
single peer's lag-\(h\) covariate inner product.  The term
\(\{\mathbb E(\bm x_{i0}'\bm x_{ih})\}^2\), by contrast, results from cross-products between
different potential peers.  When \(\mathbb E(\bm x_{i0}'\bm x_{ih})\neq0\), this creates an additional variance component
that is absent under temporal independence.
Under the short-range dependence conditions in
Assumption~\ref{ass:xweak},  $\text{Var}(S_N|\mathcal F_N)=\O_p\left(
n^2T^2+n^3T\right)$, where
the first component $n^2T^2$ is the variance scale already present under
temporal independence, and the second $n^3T$  is generated by temporal dependence when $$\sum_{h=1}^{\infty}
 \mathbb E\left\{\sigma^2\left(\bm x_{i0}\right)\sigma^2\left(\bm x_{ih}\right)\right\}
 \left\{\mathbb E\left(\bm x_{i0}'\bm x_{ih}\right)\right\}^{2}>0.$$  Their relative magnitude is governed
by
\(
{n^3T}/{(n^2T^2)}
=
 {n}/{T}\).
Thus, when \(n/T\to0\), the serial-covariance component is asymptotically
negligible; when \(n\asymp T\), the two components
may contribute at the same order; and when \(n/T\to\infty\), the
serial-covariance component may dominate
the conditional variance.  

Interestingly,
the change in the scale of the numerator is matched automatically by
the denominator.  Specifically,
\(
{\widehat V_N}/{\text{Var}(S_N|\mathcal F_N)}
\xrightarrow{p}1
\) still holds by Lemma 17. 
  Thus temporal dependence changes the
first-order scale of both the numerator and its conditional variance,
but studentization removes this scale change and leaves the limiting
distribution of the test statistic unchanged. 
\end{remark}
{
  Under temporal independence of the covariates, for every $h\ge1$, the lagged inner product has mean zero, its second moment is $\tr\left(\mathbf\Sigma_x^2\right)$, and the variance-weight product has expectation $\left(\bar\sigma^2\right)^2$. {The deterministic raw-variance scale then equals}
\(
 2n\left(n-1\right)T\left(T-1\right)
 \left(\bar\sigma^2\right)^2\tr\left(\mathbf\Sigma_x^2\right),
\)
{and the corrected conditional variance is asymptotically equivalent to this quantity, recovering the i.i.d.\ case.} 
}

\section{Numerical experiments}\label{sec:simulation}

In this section, we examine the finite-sample size and power of the proposed test.

\subsection{Design and implementation}\label{main_simu}

The data-generating process is
\begin{equation}\label{eq:sim-dgp}
 \bm y_t=\mathbf A_n\bm y_t+\bm x_t\beta+\bm\xi
          +\eta_t\bm 1_n+\bm\varepsilon_t,
 \qquad \beta=1,
\end{equation}
where $\bm x_t=(x_{1t},\ldots,x_{nt})'$,
$\xi_i\stackrel{\mathrm{i.i.d.}}{\sim}N(0,1)$, and
$\varepsilon_{it}=\sigma(x_{it})e_{it}$ with
$e_{it}\stackrel{\mathrm{i.i.d.}}{\sim}N(0,1)$. All innovation sequences and
initial draws are mutually independent. Individual effects are drawn once per
replication and held fixed over time. In the independent-covariate experiment,
$x_{it}\stackrel{\mathrm{i.i.d.}}{\sim}N(0,1)$. The time effect is either i.i.d.
$N(0,1)$ or a stationary Gaussian AR(1) process with coefficient
$\phi_\eta=0.6$ and unit marginal variance. We consider
$(n,T)=(45,20),(30,30),(20,45)$, all with $N=nT=900$.
The corresponding numbers of  instruments are
$K=n(n-1)=1980,870,380$.

We use four conditional-variance specifications:
\begin{equation}\label{eq:sim-variance-main}
 \begin{aligned}
 \mathrm{HOM}:\quad &\sigma^2(x)=1,\\
 \mathrm{HET}:\quad &\sigma^2(x)=1+0.8\tanh(x),\\
 \mathrm{HET\mbox{-}0}:\quad &\sigma^2(x)=2\mathbf 1\{x\geq0\},\\
 \mathrm{HET\mbox{-}T}:\quad
 &\sigma^2(x)=0.1+4.5\mathbf 1\{|x|>\Phi^{-1}(0.9)\},
 \end{aligned}
\end{equation}
where $\Phi(\cdot)$ is the standard-normal distribution function. Each specification
has $\mathbb E\{\sigma^2(x_{it})\}=1$ and satisfies the 
requirement in Assumption~\ref{ass:error}. 

Under the null, $\mathbf A_n=0$. Under the alternatives, we set
$\mathbf A_n=\rho\mathbf A_{0n}$ with \emph{fixed} $\rho=0.5$ for every panel
shape. The positive heterogeneous network is a directed ring with one active
link from $i+1$ to $i$, whose entry in $\mathbf A_{0n}$ is $\sqrt{3/2}$ for
odd $i$ and $\sqrt{1/2}$ for even $i$. The mixed-sign network has an entry
$1/\sqrt2$ from $i+1$ to $i$ and an entry
$(-1)^{i+1}/\sqrt2$ from $i+2$ to $i$, with indices interpreted modulo $n$.
Thus the two alternatives differ in the signs and locations of their peer
links, while both use the same fixed multiplier $\rho$. 

For comparison, we implement the method proposed by \citet{jung2026testing}, denoted by JL. Since their procedure is deveoped for the case $T>n$, JL is reported as N.A. for $(n,T)=(45,20)$ and $(n,T)=(30,30)$ and is computed
for $(n,T)=(20,45)$. For both size and power evaluations, we use 500 independent Monte Carlo replications for each configuration, and all tests are conducted at the $5\%$ nominal significance level.

\subsection{Size under independent covariates}

 Table~\ref{tab:sim-size-main} reports the empirical sizes of the proposed test and the JL test. The results show that the proposed test maintains empirical size close to the nominal significance level across all considered settings. In contrast, the JL fixed-effects test relies on homoskedasticity
for its null calibration and is unavailable in the two full-dyad
designs with $n\ge T$. When computable, it exhibits substantial
size distortion under HET-T.

\begin{table}[h!]
\centering\small
\caption{Empirical size with i.i.d. covariates (percent)}
\label{tab:sim-size-main}
\begin{tabular}{llrrrr}
\hline
 & & \multicolumn{2}{c}{i.i.d. $\eta_t$}
   & \multicolumn{2}{c}{AR(1) $\eta_t$}\\
Variance & $(n,T)$ & Proposed & JL & Proposed & JL\\
\hline
HOM   & $(45,20)$ & 6.4 & N.A. & 5.8 & N.A.\\
HOM   & $(30,30)$ & 5.6 & N.A. & 6.4 & N.A.\\
HOM   & $(20,45)$ & 6.8 & 4.6  & 6.6 & 4.0\\
HET   & $(45,20)$ & 6.6 & N.A. & 6.2 & N.A.\\
HET   & $(30,30)$ & 6.6 & N.A. & 6.2 & N.A.\\
HET   & $(20,45)$ & 6.8 & 4.2  & 6.6 & 4.2\\
HET-0 & $(45,20)$ & 6.8 & N.A. & 6.4 & N.A.\\
HET-0 & $(30,30)$ & 7.2 & N.A. & 6.2 & N.A.\\
HET-0 & $(20,45)$ & 6.2 & 4.2  & 6.6 & 4.4\\
HET-T & $(45,20)$ & 5.6 & N.A. & 5.4 & N.A.\\
HET-T & $(30,30)$ & 6.2 & N.A. & 6.0 & N.A.\\
HET-T & $(20,45)$ & 7.4 & 20.2 & 6.8 & 20.0\\
\hline
\end{tabular}
\end{table}

\subsection{Power under independent covariates}

Tables~\ref{tab:sim-power-positive} and~\ref{tab:sim-power-mixed} report the empirical rejection frequencies under the positive and mixed-sign networks, respectively, with $\rho=0.5$. The proposed test exhibits substantial power under both alternatives. For a fixed total sample size $N=nT$, its empirical power increases with $T$ across the considered panel configurations. Moreover, the proposed test generally achieves higher rejection frequencies than the JL test in settings where both tests are applicable.

\begin{table}[h!]
\centering\small
\caption{Positive heterogeneous network: empirical rejection frequencies (percent)}
\label{tab:sim-power-positive}
\begin{tabular}{llrrrr}
\hline
 & & \multicolumn{2}{c}{i.i.d. $\eta_t$} & \multicolumn{2}{c}{AR(1) $\eta_t$}\\
Variance & $(n,T)$ & Proposed & JL & Proposed & JL\\
\hline
HOM   & $(45,20)$ & 75.6 & N.A. & 75.0 & N.A.\\
HOM   & $(30,30)$ & 93.8 & N.A. & 95.4 & N.A.\\
HOM   & $(20,45)$ & 99.8 & 98.4 & 99.6 & 97.2\\
HET   & $(45,20)$ & 74.4 & N.A. & 75.0 & N.A.\\
HET   & $(30,30)$ & 96.8 & N.A. & 93.0 & N.A.\\
HET   & $(20,45)$ & 100.0 & 97.2 & 99.6 & 99.0\\
HET-0 & $(45,20)$ & 73.8 & N.A. & 75.2 & N.A.\\
HET-0 & $(30,30)$ & 93.8 & N.A. & 94.2 & N.A.\\
HET-0 & $(20,45)$ & 99.8 & 97.2 & 100.0 & 97.2\\
HET-T & $(45,20)$ & 77.0 & N.A. & 75.4 & N.A.\\
HET-T & $(30,30)$ & 94.6 & N.A. & 95.0 & N.A.\\
HET-T & $(20,45)$ & 100.0 & 99.2 & 99.4 & 98.6\\
\hline
\end{tabular}
\end{table}

\begin{table}[h!]
\centering\small
\caption{Mixed-sign network: empirical rejection frequencies (percent)}
\label{tab:sim-power-mixed}
\begin{tabular}{llrrrr}
\hline
 & & \multicolumn{2}{c}{i.i.d. $\eta_t$} & \multicolumn{2}{c}{AR(1) $\eta_t$}\\
Variance & $(n,T)$ & Proposed & JL & Proposed & JL\\
\hline
HOM   & $(45,20)$ & 64.8 & N.A. & 65.6 & N.A.\\
HOM   & $(30,30)$ & 87.6 & N.A. & 89.8 & N.A.\\
HOM   & $(20,45)$ & 99.0 & 92.8 & 98.8 & 93.0\\
HET   & $(45,20)$ & 65.2 & N.A. & 66.0 & N.A.\\
HET   & $(30,30)$ & 87.4 & N.A. & 89.4 & N.A.\\
HET   & $(20,45)$ & 97.8 & 91.8 & 98.6 & 92.8\\
HET-0 & $(45,20)$ & 63.4 & N.A. & 63.6 & N.A.\\
HET-0 & $(30,30)$ & 90.6 & N.A. & 89.0 & N.A.\\
HET-0 & $(20,45)$ & 98.4 & 90.4 & 98.8 & 93.8\\
HET-T & $(45,20)$ & 65.6 & N.A. & 63.8 & N.A.\\
HET-T & $(30,30)$ & 83.4 & N.A. & 89.4 & N.A.\\
HET-T & $(20,45)$ & 98.6 & 97.2 & 98.8 & 97.8\\
\hline
\end{tabular}
\end{table}

\subsection{Weakly dependent covariates}

We replace i.i.d. covariates by independent trajectories across individuals,
each following the stationary Gaussian process
\begin{equation}\label{eq:sim-x-ar1}
 x_{it}=0.6x_{i,t-1}+\sqrt{1-0.6^2}\,v_{it},\qquad
 v_{it}\stackrel{\mathrm{i.i.d.}}{\sim}N(0,1),\quad x_{i1}\sim N(0,1).
\end{equation}
All other features of the design are the same as in Section \ref{main_simu}, including the four variance
specifications in \eqref{eq:sim-variance-main} and fixed $\rho=0.5$ under
both network alternatives.  
 
Table~\ref{tab:sim-x-ar1} reports the empirical sizes under the null and the empirical powers under the alternatives. The results indicate that the proposed test continues to exhibit reasonable size control and substantial power in the presence of temporal dependence in the covariates.  

\begin{table}[h!]
\centering\small
\caption{AR(1) covariates:  empirical rejection frequencies (percent)}
\label{tab:sim-x-ar1}
\begin{tabular}{llrrrrrr}
\hline
 & & \multicolumn{3}{c}{i.i.d. $\eta_t$} & \multicolumn{3}{c}{AR(1) $\eta_t$}\\
Variance & $(n,T)$ & Null & Positive & Mixed & Null & Positive & Mixed\\
\hline
HOM   & $(45,20)$ & 6.4 & 94.4 & 86.6 & 5.2 & 94.6 & 97.8\\
HOM   & $(30,30)$ & 5.6 & 98.4 & 94.6 & 6.8 & 99.4 & 99.4\\
HOM   & $(20,45)$ & 6.8 & 99.6 & 98.8 & 8.8 & 100.0 & 99.8\\
HET   & $(45,20)$ & 7.2 & 92.8 & 83.6 & 8.2 & 95.2 & 97.0\\
HET   & $(30,30)$ & 8.6 & 97.6 & 95.8 & 7.4 & 98.2 & 99.8\\
HET   & $(20,45)$ & 7.2 & 99.8 & 99.4 & 8.0 & 100.0 & 100.0\\
HET-0 & $(45,20)$ & 7.0 & 95.2 & 88.4 & 7.2 & 95.2 & 97.0\\
HET-0 & $(30,30)$ & 6.6 & 97.8 & 92.6 & 5.6 & 98.4 & 98.8\\
HET-0 & $(20,45)$ & 7.0 & 99.8 & 99.0 & 5.6 & 99.8 & 100.0\\
HET-T & $(45,20)$ & 8.0 & 91.8 & 83.0 & 6.0 & 93.8 & 97.6\\
HET-T & $(30,30)$ & 6.4 & 95.8 & 92.6 & 7.8 & 97.2 & 98.2\\
HET-T & $(20,45)$ & 8.4 & 99.8 & 98.8 & 7.4 & 99.8 & 100.0\\
\hline
\end{tabular}
\end{table}

\section{Application}
\label{sec:application}

In this section, we apply the proposed test to state tax competition and growth spillovers
among OECD countries. In both applications, we include individual and time fixed
effects and test whether all peer effects
are zero. Every other individual is treated as a potential peer, with no restrictions 
on the geographical or economic structure of the network.

\subsection{State tax competition}
\label{subsec:state-tax-competition}

Under yardstick competition, voters compare tax policies across jurisdictions,
giving state governments an incentive to respond to tax changes elsewhere
\citep{besleycase1995yardstick}. \citet{depaula2025network} study these
interactions in a panel model with an unobserved network. We use their
replication data to test the joint absence of outcome interactions in the
balanced panel of 48 contiguous U.S. states over 1962--1988, with
$n=48$, $T=27$, and $N=1{,}296$. Since $n>T$, the test of \citet{jung2026testing} is inapplicable in this setting.

Following the source specification, let
$\Delta_2 a_{it}=a_{it}-a_{i,t-2}$. The response $y_{it}$ is the two-year
change in total state taxes per capita, rescaled by $10^3$.
\citet{depaula2025network} use two-year changes because implementing a
tax program may take a governor more than one year. The five own-state
covariates are two-year changes in income per capita, the unemployment rate,
and the young and elderly population shares, together with the governor's
age in levels.

We specify
\begin{equation}
    y_{it}
    =\sum_{j\ne i}\alpha_{ij}y_{jt}
     +x_{it}'\beta+\xi_i+\eta_t+\varepsilon_{it},
    \qquad i=1,\ldots,48,\quad t=1,\ldots,27.
    \label{eq:tax-application}
\end{equation}
State effects $\xi_i$ account for persistent differences in fiscal
institutions and policy preferences, and year effects $\eta_t$ absorb
additive national shocks. The coefficient $\alpha_{ij}$ measures the
response to state $j$'s tax change in state $i$'s structural equation.
These coefficients are constant over the sample period but may differ
across ordered state pairs.  

The null hypothesis is
\[
    H_0^{\mathrm{tax}}:\quad
    \alpha_{ij}=0\quad\text{for all }i\ne j.
\]
Using all five covariates of each potential peer gives
$K=48\times47\times5=11{,}280$ candidate instrument columns. 
The test yields $T_N=4.9303$ and
$p_{\mathrm N}=4.10\times10^{-7}$, rejecting
$H_0^{\mathrm{tax}}$ at the $1\%$ significance level, providing   evidence of interactions in state tax adjustments.

\subsection{Growth spillovers among OECD countries}
\label{subsec:oecd-growth}

Knowledge diffusion and technological spillovers generate international
productivity interdependence in the spatially augmented Solow model of
\citet{erturkoch2007growth}. We follow the simplified panel specification
used by \citet{jung2026testing} to test for growth spillovers among OECD
countries, retaining the two peer covariates separately in the instrument
dictionary.

We reconstruct the balanced panel of the same 28 countries over 1975--2015
from Penn World Table 10.0 and OECD working-age population data.
The sample has $n=28$, $T=41$, and $N=1{,}148$.
Let $g_{it}$ denote output-side real GDP per worker and set
$y_{it}=\log g_{it}$.
Let $p_{it}$ denote the five-year average annual growth rate of the
population aged 15--64, and let $s_{it}$ denote the average investment
share in GDP over $t-5,\ldots,t-1$.
The own-country regressors are
    $x_{it,1}=\log(p_{it}+0.05)$,
  $x_{it,2}=\log s_{it}$, 
where $0.05$ is the assumed sum of the rates of technological progress
and capital depreciation.


The model is
\begin{equation}
    y_{it}
    =\sum_{j\ne i}\alpha_{ij}y_{jt}
     +\beta_1\log(p_{it}+0.05)+\beta_2\log s_{it}
     +\xi_i+\eta_t+\varepsilon_{it},
    \label{eq:oecd-application}
\end{equation}
where $i=1,\ldots,28$ and $t=1975,\ldots,2015$.
Country and year effects account for persistent country differences and
additive global shocks with common loadings. The response is log productivity
in levels, and the interaction coefficients may vary across ordered country
pairs but are constant over time. As in the state tax specification, we
exclude direct effects of other countries' regressors conditional on
own-country controls and fixed effects.

We test
\[
    H_0^{\mathrm{OECD}}:\quad
    \alpha_{ij}=0\quad\text{for all }i\ne j.
\]
Retaining both peer covariates gives
$K=28\times27\times2=1{,}512$ candidate instrument columns.
The proposed test yields $T_N=43.1869$, with a standard-normal reference $p$-value of $p_{\mathrm N}<0.001$. We therefore reject $H_0^{\mathrm{OECD}}$, providing evidence of cross-country outcome interactions.

As a benchmark, \citet{jung2026testing} combine the peer
regressors as $x_{jt,1}+x_{jt,2}$, giving $28\times27=756$
candidate peer-instrument columns. They report a near-zero $p$-value;
their chi-square-approximation statistic
$C_{\mathrm{JL}}=1070.70$ and the $1\%$ critical value $810.22$.
Thus, both analyses reject the  null at the $1\%$ level.

Table~\ref{tab:oecd-sensitivity} reports results under alternative covariate
scaling, investment windows, and sample periods. Rejection at the $1\%$ significance
level is retained after separately standardizing the two log covariates,
including the current year in the five-year investment window, and
restricting the sample to either 1975--1994 or 1995--2015.
The two subperiods have $T=20$ and $T=21$, respectively, both below $n=28$.

\begin{table}[h!]
    \centering
    \caption{Tests for peer effects in the two applications}
    \label{tab:empirical-peer-tests}
    \small
    \begin{tabular}{lrrl}
        \toprule
        Application and method & Peer columns & Statistic & Reference $p$ \\
        \midrule
        State taxes: proposed
            & $11{,}280$ & $T_N=4.9303$ & $4.10\times10^{-7}$ \\
        OECD: proposed, full dictionary
            & $1{,}512$ & $T_N=43.1869$ & $<0.001$ \\
        OECD: Jung--Liu, published
            & $756$ & $C_{\mathrm{JL}}=1070.70$ & Near zero \\
        \bottomrule
    \end{tabular}
\end{table}

\begin{table}[h!]
    \centering
    \caption{Sensitivity of the OECD application}
    \label{tab:oecd-sensitivity}
    \begin{tabular}{lrr}
        \toprule
        Specification & $T_N$ & $p_{\mathrm N}$ \\
        \midrule
        Original log units, 1975--2015 & $43.1869$ & $<0.001$ \\
        Separately standardized log covariates & $44.5103$ & $<0.001$ \\
        Investment window includes current year & $42.3302$ & $<0.001$ \\
        Original log units, 1975--1994 & $20.8546$ & $<0.001$ \\
        Original log units, 1995--2015 & $19.1547$ & $<0.001$ \\
        \bottomrule
    \end{tabular}
    \par\smallskip
    \begin{minipage}{0.97\linewidth}
        \footnotesize
        Notes: Standardization
        changes the relative weights of the covariates in the Gram matrix.
        The remaining specifications use the original log units.
    \end{minipage}
\end{table}

\section{Concluding remarks}\label{sec:conclusion}
{
This paper develops a test for peer effects in linear panels with an unspecified, time-invariant interaction network and two-way fixed effects. The test uses the full set of candidate peer covariates but does not invert their high-dimensional Gram matrix. Its essential adjustment is to center the effective quadratic form in error coordinates, rather than merely deleting a diagonal in estimated-residual coordinates. This preserves exact conditional centering under the maintained heteroskedastic error model and accommodates the dependence created by fixed-effects removal and restricted ordinary least squares estimation.  
The null approximation requires both panel dimensions to diverge but does not prescribe their relative rates. A residual-based variance estimator provides feasible studentization, including when the conditional variance vanishes on part of the covariate support. {We also establish consistency when the peer-effect signal is sufficiently separated from the null.} The extension to temporally dependent covariates shows that a change in the stochastic scale need not require a change in the test: the realized kernel and the same variance estimator incorporate the additional covariate dependence.


Several directions remain for future research. {First, extending the consistency analysis to temporally dependent covariates is a further theoretical question.} 
Second, under a sparse network, a max-type statistic may offer higher power by targeting a small number of large peer effects. Developing procedures that adapt to both dense and sparse alternatives is a natural extension. Another direction concerns post-rejection analysis. {The present paper tests the joint null of no peer effects, but does not identify which coefficients are nonzero once the null is rejected.} Recovering and conducting inference on the active peer effects in an unknown network is an important next step. Finally, allowing the peer-effect coefficients to vary over time, \(\alpha_{ijt}\), would accommodate evolving interaction patterns and extend the framework beyond time-invariant networks.
}

\bibliographystyle{apalike}
\bibliography{ref}
 
\end{document}